\documentclass[showpacs,aps,twocolumn]{revtex4}

\usepackage{graphicx}
\begin{document}

\title{Persistent spin current in spin-orbit coupling systems in the absence
of an external magnetic field}
\author{Qing-feng Sun$^{1,2,\ast}$ and X. C. Xie$^{2,3}$}
\affiliation{$^1$Beijing National Lab for Condensed Matter Physics
and Institute of Physics, Chinese Academy of Sciences, Beijing
100080, China \\
$^2$Department of Physics, Oklahoma State University, Stillwater,
Oklahoma 74078 \\
$^3$International Center for Quantum structures, Chinese Academy
of Sciences, Beijing 100080, China }
\date{\today}

\begin{abstract}
The spin-orbit coupling systems with a zero magnetic field is studied
under the equilibrium situation, {\it i.e.}, without a voltage bias.
A persistent spin current is
predicted to exist under most circumstances, although the persistent
charge current and the spin accumulation are identically zero. In particular, a
two-dimensional quantum wire is investigated in
detail. Surprisingly, a persistent spin current is found to
flow along the confined direction, due to
the spin precession in accompany with the particle motion. This provides
an interesting example of constant spin flowing without
inducing a spin accumulation, contrary to common intuition.
\end{abstract}


\maketitle

In last two decades, the persistent charge currents in mesoscopic
rings threaded by a magnetic flux have been extensively
investigated.\cite{ref1,ref2,ref3} The persistent current is a
pure quantum effect appeared in the equilibrium situation. In an
ideal case it can last forever without dissipation. In recent
years, another subject, the spin-orbit (SO) interaction in
semiconductors, has also attracted a great deal of
attention.\cite{ref4} The SO coupling is an intrinsic interaction,
originated from the relativistic effect. By using the SO coupling,
the electron spins can be conveniently manipulated and controlled
experimentally, this may have large potential applications in new
generation of spin-based electronic devices.\cite{ref4} Due to the
SO coupling, a moving spin is under an equivalent magnetic field
with the field strength depending on the momentum of the particle.
Thus, it is nature to ask whether there exists a persistent
current in a SO coupling system, similar as for a system with an
external magnetic field. Some recent works have begun to explore
in this direction.\cite{ref6,Rashba,nref1}

In this Letter, our goal is to systematically address the
question: Is there a persistent spin or charge current in a SO
coupling system without an external electric or a magnetic field?
Furthermore, we focus on the physical properties of the persistent
current if they do exist. Before studying a concrete system, we
first provide a general discussion from symmetry point of view.
First, consider a system without a magnetic field nor the SO
coupling. Thus, the Hamilton is $H=\frac{\vec{p}^2}{2m} +V({\bf
r})$, with $\vec{p}$ being the momentum operator and $V({\bf r})$
the potential energy. This Hamilton contains two symmetries: (i)
The time-reversal invariance, i.e. $[T,H]=0$ with $T$ being the
time-reversal operator.\cite{ref5} This symmetry leads to the
Kramer degeneracy: i.e. if $\Psi({\bf r})$ is an eigenstate of
$H$, the state $T\Psi({\bf r})$ is also an eigenstate with the
same energy and $<\Psi|T\Psi>=0$.\cite{ref5} (ii) The Hamilton
commutes with the three Pauli matrices $\sigma_i$ ($i=x,y,z$),
i.e. $[\sigma_i, H]=0$. This symmetry leads to the degeneracy:
$\Psi = (\varphi({\bf r}),0)^T$ and $\Psi' =(0,\varphi({\bf
r}))^T$ have the same eigenenergy and $<\Psi|\Psi'>=0$. In the
equilibrium case with the degenerated energy eigenstates having
the same occupational probability, it is known that the persistent
charge current is identically zero because of the time-reversal
symmetry (i) and so is the persistent spin current because of the
symmetry (ii). Second, consider a mesoscopic ring threaded by a
magnetic flux $\Phi$. In this system, the time-reversal symmetry
(i) is broken but the symmetry (ii) survives. Then the persistent
spin current must vanish because of the symmetry (ii), but the
persistent charge current is in general nonzero. In fact, this
kind of systems has been extensively investigated and is well
understood by now.\cite{ref1,ref2,ref3} Finally, consider a system
having SO coupling but without a magnetic field. The Hamilton is
$H=\frac{\vec{p}^2}{2m} +V({\bf r}) +H_{so}$. In the present
system, the symmetry (ii) is destroyed, but the time-reversal
symmetry still holds since $[T,H_{so}]=0$.\cite{ref5} Hence the
Kramer degeneracy of $\Psi$ and $T\Psi$ exists. These two states
have exactly opposite charge current densities but identical spin
current densities. So the persistent charge current and the local
spin accumulation must be zero everywhere in equilibrium. However,
the persistent spin current usually exists because of the absence
of the symmetry (ii).

Next we consider a concrete system having a SO coupling. We focus
on whether the spin current is indeed nonzero in equilibrium, as
well as study its physical property if it does exist. Our system
is a two-dimensional electron gas (2DEG)  with the Rashba SO
interaction, and its Hamilton is:
\begin{equation}
  H=\frac{p_x^2 +p_y^2}{2m} +V(x,y)
    + \frac{\alpha}{\hbar}\bullet (\hat{\sigma} \times \vec{p}).
\end{equation}
The Rashba SO interaction originates from an asymmetrical
interface electric field in the z-direction,\cite{ref7,ref8} and
its strength $\alpha$ can be varied experimentally.\cite{nref2}
The potential energy $V(x,y)=0$ while $0\leq y \leq L$ and
$V(x,y)=\infty$ otherwise, i.e. we consider a quasi
one-dimensional quantum wire as shown in Fig.1a. The present
Hamiltonian can be exactly solved. Due to the fact that
$[p_x,H]=0$, $k_x$ is still a good quantum number. For a fixed
$k_x$, the general solutions of the differential equation
$H\Psi(x,y) = E\Psi(x,y)$ are:
\begin{equation}
  \Psi(x,y) =\sum_{i=\{1,2,3,4\}} a_i \Psi_i(x,y),
\end{equation}
where $a_i$ are constants to be determined by boundary
conditions, and $\Psi_i(x,y)$ are:
\begin{eqnarray}
 \Psi_1(x,y) & =& \frac{\sqrt{2}}{2} e^{ik_x x} e^{i k_y^+ y}
    \left( \begin{array}{c} -(k_y^+ +ik_x)/k_+ \\ 1 \end{array}
    \right), \nonumber \\
 \Psi_2(x,y) & =& \frac{\sqrt{2}}{2} e^{ik_x x} e^{-i k_y^+ y}
    \left( \begin{array}{c} (k_y^+ - ik_x)/k_+ \\ 1 \end{array}
    \right), \nonumber \\
 \Psi_3(x,y) & =& \frac{\sqrt{2}}{2} e^{ik_x x} e^{i k_y^- y}
    \left( \begin{array}{c} (k_y^- +ik_x)/k_- \\ 1 \end{array}
    \right), \nonumber \\
 \Psi_4(x,y) & =& \frac{\sqrt{2}}{2} e^{ik_x x} e^{-i k_y^- y}
    \left( \begin{array}{c} -(k_y^- - ik_x)/k_- \\ 1 \end{array}
    \right). \nonumber
\end{eqnarray}
Here $k_y^{\pm} \equiv \sqrt{k_{\pm}^2 - k_x^2}$, $k_{\pm}
=\sqrt{\frac{2m}{\hbar^2} E +k_R^2}\pm k_R$, and $k_R =\alpha
m/\hbar^2$. Using the boundary conditions $\Psi(x,0/L)=0$,
the eigenenergy $E$ can be solved. Fig.2 shows $E$ versus $k_x$
and it exhibits a series of sub-band structures due to the
confinement in the transverse direction. For a narrow quantum wire
(e.g. $L=50nm$, see Fig.2a), the sub-bands are well apart and separated
from each other. The curves of $E$ versus $k_x$ are of
parabolic shape. On the other hand, for a wide quantum wire with
$L=200nm$ (see Fig.2b), the sub-bands are very dense, and there
are obvious mixtures between different sub-bands. These results
are similar as in the previous studies with a parabolic
confinement potential.\cite{ref12}

For given eigenenergy $E$ and $k_x$, {\it i.e.}, corresponding to
each point on the curves in Fig.2, the eigen wave-function
$\Psi(x,y)$ (i.e. the coefficient $a_i$ in Eq.(2)) can be obtained
using the boundary conditions and the normalization equation. For
convenience, we take the state density normalization:
$<\Psi|\Psi>=1/(dE/dk) \propto \rho(E)$. Then the (linear) spin
current density ${\bf j}_s(x,y)$ (${\bf j}_s \equiv Re
\Psi^{\dagger}\vec{v}\hat{\vec{s}}\Psi$), the angular spin current
density (or spin torque) $\vec{j}_{\omega}(x,y)$
($\vec{j}_{\omega} \equiv Re \Psi^{\dagger} (d\hat{\vec{s}}/dt)
\Psi$), and the local spin density $\vec{s}(x,y)$ ($\vec{s}=
\Psi^{\dagger}\hat{\vec{s}}\Psi$) for every state $\Psi(x,y)$ can
also be calculated. Here ${\bf j}_s$ and $\vec{j}_{\omega}$
describe respectively the translational and the rotational motion
of a spin, and they satisfy the spin continuity equation
$\frac{d}{dt} \vec{s} = -\nabla \bullet {\bf j}_s +
\vec{j}_{\omega}$ (see \cite{ref10} for a detail discussion). As
the last step, summing over all the occupied states, the
persistent spin current density and the local spin density in the
equilibrium can be obtained straightforwardly: e.g. ${\bf
j}_s^T(x,y) =\int dE \sum_n f(E){\bf j}_{s n}(x,y,E)$ where $f(E)$
is the Fermi distribution function and $n$ is the sub-band index.

Next we numerically investigate the linear and angular persistent
spin current densities ${\bf j}^T_s$ and $\vec{j}^T_{\omega}$. It
clearly shows that ${\bf j}^T_s$ and $\vec{j}^T_{\omega}$ indeed
exist in (Rashba) SO coupling system in the absence of a magnetic field
(See Fig.3). We mention again that in the present system the
persistent charge current and the spin accumulation are identically
zero because of the time-reversal symmetry. Our numerical
calculations also confirmed this result. The non-zero elements of
${\bf j}^T_s$ and $\vec{j}^T_{\omega}$ are $j_{s,yx}^T$,
$j_{s,xy}^T$, $j_{s,xz}^T$, and $j_{\omega x}^T$. Those non-zero
elements are all independent of $x$ because of the translational
invariance along the $x$ direction. Fig.3 shows $j_{s,yx}^T$,
$j_{s,xy}^T$, and $j_{s,xz}^T$ versus $y$, in which $j_{s,yx}^T$
and $j_{s,xy}^T$ are symmetric and $j_{s,xz}^T$ are
anti-symmetric. $j_{\omega x}^T$ is not shown here because it is
related to $j_{s,yx}^T$ through the spin continuity equation
for a steady state:
$j_{\omega x}^T = \frac{d}{dy} j_{s,yx}^T$. Surprisingly,
$j_{s,yx}^T$ turns out to be non-zero. This means that there
exists a persistent spin current flowing along the $y$-direction
(confined direction) with the spin pointing in the $x$-direction.
This is unexpected since common belief is that if there is spin
current in the confined direction, there will be a spin
accumulation at sample boundaries. However, in this system there
is no spin accumulation anywhere, but a spin current continuously
and persistently flows along the $y$-direction. In the following,
we provide an understanding why this is possible.

The upshot of the story is that because of the role of the angular
spin current $\vec{j}^T_{\omega}$ in the spin continuity equation,
the spin will precess in accompany with the particle motion. As a
consequence, the spin current can flow in the confined direction
without causing spin accumulation.

In order to understand the origin of the non-zero $j_{s,yx}^T$
in more detail, we
investigate the spin current density of a single state, namely, the
state A in Fig.2a. Here the non-zero elements of ${\bf
j}_s$ and $\vec{j}_{\omega}$ are still $j_{s,yx}$, $j_{s,xy}$,
$j_{s,xz}$, and $j_{\omega x}$. The others elements as well as $s_x$
are zero for all ($x,y$). Fig.4a,b show $j_{s,yx}$, $j_{\omega x}$,
$s_y$ and $s_z$ of the state A. The curves for $j_{s,xy}$ and
$j_{s,xz}$ are similar to $s_y$ and $s_z$, respectively, so they
are not shown here. It is surprising that there is a
non-zero spin current flowing along the $y$-axis with spin pointing to
the $x$-direction, but the element $s_x$ is exactly zero everywhere.

Before to explain the spin current density, let us first study
what is the electron motion in the state A. Without the
Rashba coupling, {\it i.e.}, $k_R \rightarrow 0$,
its wave-function is $\Psi(x,y) = N e^{ik_x x} \sin (\pi y/L)
(i,1)^T = \frac{N}{2i} e^{ik_x x} (e^{i\pi y/L}-e^{-i\pi y/L})
(i,1)^T \equiv \Psi_+ +\Psi_-$. This is a state with electron spin
towards $-y$ direction, propagating along the $x$ axis while
reflecting back and forth in the transverse $y$-direction, as
shown in Fig.1a. Here $\Psi_+$ and $\Psi_-$ represent the
traveling waves along $+y$ and $-y$ directions, respectively.
While $k_R \not= 0$, the wave-function $\Psi$ can still be
separated into $\Psi_+$ and $\Psi_-$, with $\Psi_+ =a_1\Psi_1+
a_3\Psi_3$ and $\Psi_- =a_2\Psi_2+ a_4\Psi_4$. So the electron
motion looks still like the case of $k_R=0$. However, its spin is
no longer fixed at $-y$ direction anymore, and the electron spin
precesses while an electron makes its motion because of the
existence of the Rashba SO coupling.\cite{ref13} Let us describe
the spin precession in detail for the state A. Assume an electron
at the beginning is on the edge of $y=0$ moving in the $+y$
direction (see the point 1 in Fig.1a). This state is described by
wave-function $\Psi_+(x,0)$. The spin of $\Psi_+(x,0)$ is mainly
pointing to $-y$ with a slight tilt towards $+z$ (see the arrow 1
in Fig.1b). When an electron makes the motion from $1\rightarrow 2
\rightarrow 3$ (described by $\Psi_+(x,y)$), then $3\rightarrow 4
\rightarrow 5$ (described by $\Psi_{-}(x,y)$), the spin vector
makes the rotation as shown in Fig.1b. When the electron comes
back to the original side, the spin also precesses back to the
original direction it started with. The above process repeats
itself.

With the above knowledge, we are ready to explain the origin of the
spin current. We still use the state A as an example. Note that
everywhere inside the sample one can split the wave-function:
$\Psi(x,y) = \Psi_+(x,y) + \Psi_-(x,y) $. For the $+y$-direction
moving wave $\Psi_+$, its spin element $s_x$ is negative (e.g. see
the arrow 2 in Fig.1b), so it induces a negative $j_{s,yx}$. For
the $-y$-direction moving wave $\Psi_-$, its spin element $s_x$ is
positive (e.g. see the arrow 4 in Fig.1b), so it also has a
negative $j_{s,yx}$ because the motion is along $-y$. Therefore, a
negative $j_{s,yx}$ emerges for the state A (see Fig.4a). The
behavior of other elements of ${\bf j}_s(x,y)$,
$\vec{j}_{\omega}(x,y)$, and $\vec{s}(x,y)$ can also be explained
in the same manner. Here we emphasize two points: (i) The
behaviors of all elements of ${\bf j}_s$, $\vec{j}_{\omega}$, and
$\vec{s}$ can be understood with the above picture of the electron
motion accompanied by its spin precession. (ii) Due to the fact
that the motion is periodic and repeated, this motion does not
change the local spin density, and the spin current can
persistently exist.

In the above analysis, the electron is in the lowest sub-band. The
electron motion in a higher sub-band is similar, only its
spin precession is in a more complicated manner.
The electron  motion plus the spin procession can
also generate nonzero $j_{s,yx}$, $j_{s,xy}$, $j_{s,xz}$, and
$j_{\omega x}$. Fig.4 (c) and (d) show $j_{s,yx}$ and $j_{\omega
x}$ for the state E (see Fig.2a) in the 2nd sub-band and the state
F (see Fig.2b) in the 4th sub-band, respectively. We emphasize
that those results and the nature of the electron motion show that the
wave-functions are extended in the transverse direction
and they do not form edge states.
This is quite different from the case with a strong
magnetic field, in which edge states appear.

The state B (see Fig.2a), the other state in the 1st sub-band,
describes a similar electron motion (shown in Fig.1a,c), but the
spin is mainly towards $+y$ axis and it tilts to $-z$ axis at
beginning (point 1). So that its spin current elements are
opposite in sign with those of the state A, including $j_{s,yx}$
and $j_{\omega x}$ which are shown in Fig.5a,b. Thus, they
partially cancel each other out, but the remaining net spin
currents are still quite large. Also note that there are total
four degenerated states (A, B, C, and D) at $E=0.005$ and
$L=50nm$. The other two states C and D are the time-reversal
states of the states A and B, respectively. Therefore, the spin
currents of C and D are exactly the same with those  of A and B,
but the local spin densities and charge currents are exact
opposite with those of A and B. So that the total spin current of
those four states doubles that of A and B, and the local spin
density or the charge current vanishes. While at a higher $E$ or
for a wider sample width $L$, more sub-bands are involved and more
degenerated states emerge. Fig.5c shows the spin currents
$j_{s,yx}$ and $j_{\omega x}$ for $E=0.005$ and $L=200nm$ with 4
active sub-bands. The total spin current of those states still
remain a large value.

Integrating over energy $E$, the persistent spin currents are
obtained, which are shown in Fig.3. From the above analysis, the
spin current $j_{s, yx}$ flowing in the confined direction can
indeed exist constantly and it does not induce any spin
accumulation. This is due to the spin precession during the
electron motion, which makes a nonzero angular spin current
$j_{\omega x}$. Finally, we make several remarks. (i) All electrons
in the system, not only the ones near the Fermi surface,
contribute to the persistent spin current. This is quite different
from a voltage-biased system, in which only the electrons near the
Fermi surface contribute to a current. (ii) It is a universal
feature that a persistent spin current appears in the (Rashba) SO
coupling systems. For example, in a mesoscopic ring or a disc
device, ${\bf j}_s$ and $\vec{j}_{\omega}$ are usually nonzero.
Moreover, impurities in the device will not destroy the persistent
spin current, the spin flow will only be modified by the impurity
potentials. (iii) The persistent spin current can induce an
electric field, which offer a way of detecting the persistent spin
current.\cite{ref10,ref14}

In summary, we predict that a persistent spin current will
commonly appear in a spin-orbit coupling sample in the absence of
a magnetic field under zero voltage bias, although the spin
accumulation and the persistent charge current are identically zero.
Moreover, a spin current ${\bf j}_s$ is found to flow along the
confined direction in the Rashba SO coupling two-dimension quantum
wires. This implies that a spin current ${\bf j}_s$ does not
necessarily related to spin accumulation.

{\bf Acknowledgments:} We gratefully acknowledge financial support
from the Chinese Academy of Sciences and NSFC under Grant No.
90303016 and No. 10474125. XCX is supported by US-DOE under Grant
No. DE-FG02-04ER46124 and NSF-MRSEC under DMR-0080054.

\newpage
\begin{figure}

\caption{(color online) (a) Schematic diagram for a two-dimensional
quantum wire device. The arrows describe the direction of
particle motion for a state with a positive $k_x$.
(b) and (c) are
schematic diagrams for the spin precessions in the states A and B in
Fig.2, respectively. }\label{fig:1}

\caption{(color online) The dispersion relation, $E$ vs. $k_x$,
for $L=50nm$ (a) and $200nm$ (b) with $k_R=0.01/nm$. }
\label{fig:2}

\caption{(color online) $j_{s,xy}^T$ (blue dashed line),
$j_{s,xz}^T$ (red dotted line), and $j_{s,yx}^T$ (black solid
line) vs $y$ for $L=50nm$ and $E_F=0.02$ (a), and $L=200nm$ and
$E_F=0.005$ (b), respectively. The other parameters are $k_R=0.01/nm$
and the temperature ${\cal T}=0$. Here $\rho=2m E_F/\hbar^2$ is
the electron area density.
 } \label{fig:3}

\caption{ (a) and (b) show $j_{s,yx}$ (solid line in (a)), $j_{\omega
x}$ (dotted line in (a)), $s_{y}$ (solid line in (b)), and $s_{z}$
(dotted line in (b)) vs. $y$ for the state A in Fig.2a. (c) and
(d) show $j_{s,yx}$ (solid line) and $j_{\omega x}$ (dotted line)
vs. $y$ for the state E in Fig.2a and the state F in Fig.2b,
respectively. } \label{fig:4}

\caption{(color online) Show $j_{s,yx}$ (a) and $j_{\omega
x}$ (b) vs. $y$ for the state A (black solid line), the state B
(black dotted line), and the sum of the two (red solid line). (c) Show the
total $j_{s,yx}$ (black solid line) and $j_{\omega x}$ (red dotted
line) of the 16 states for $E=0.005$ and $L=200nm$ (see Fig.2b) vs.
$y$.
 } \label{fig:5}

\end{figure}

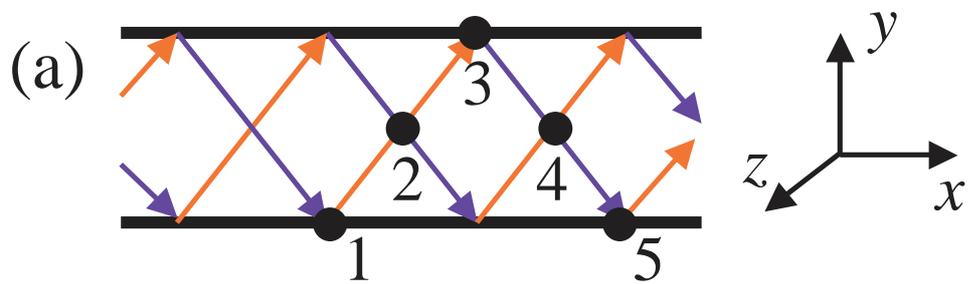
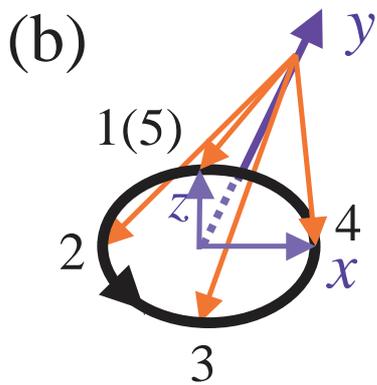
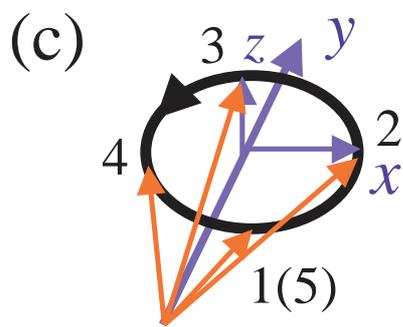
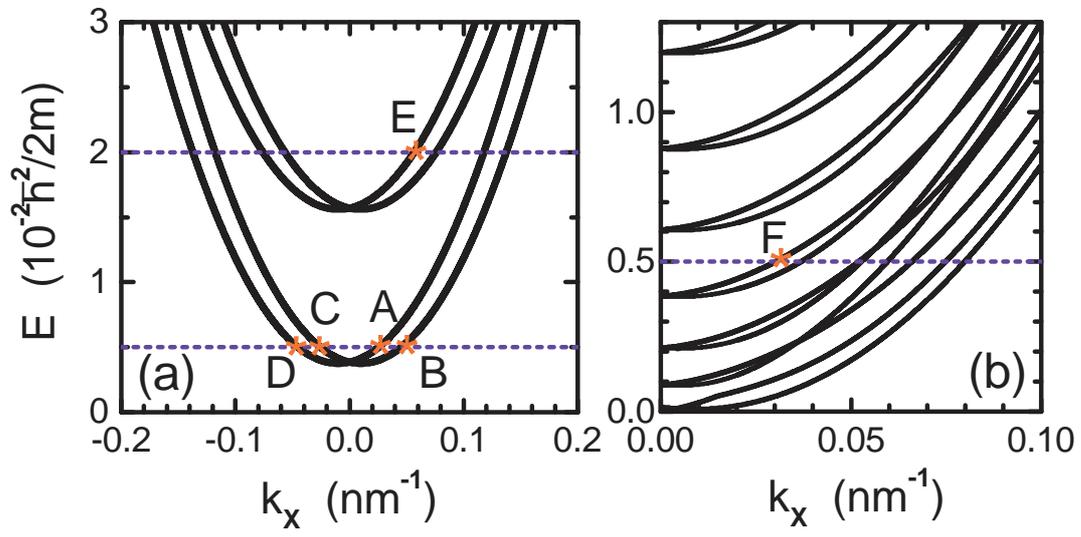
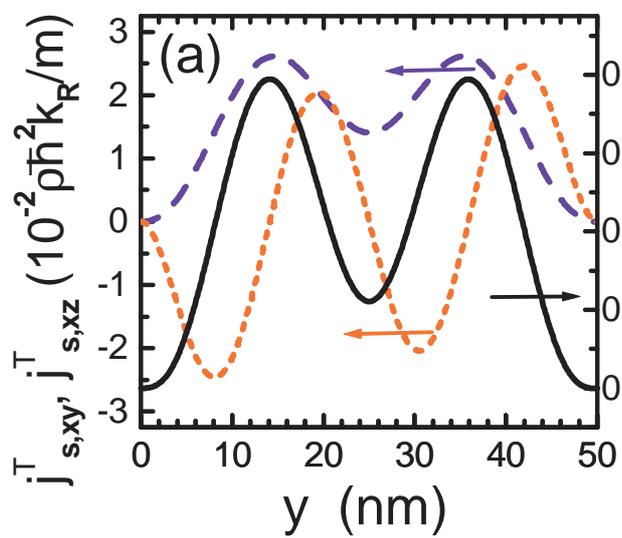
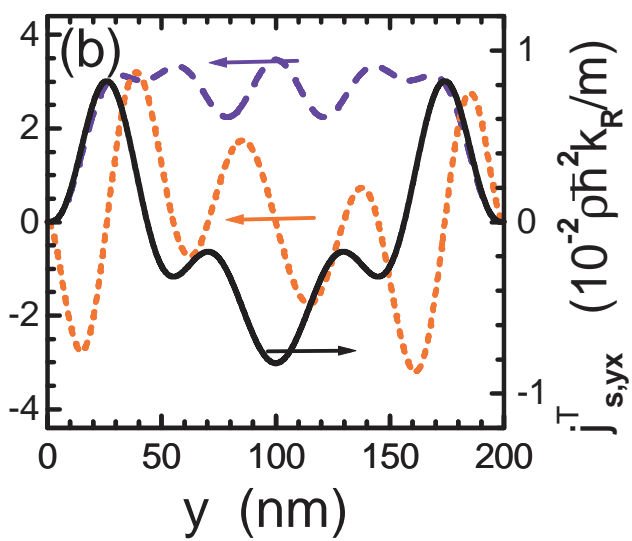
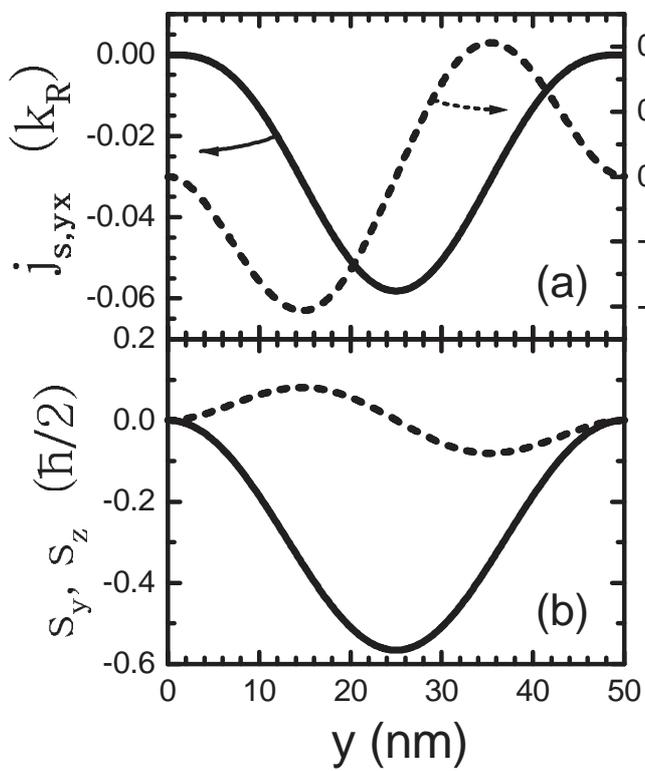
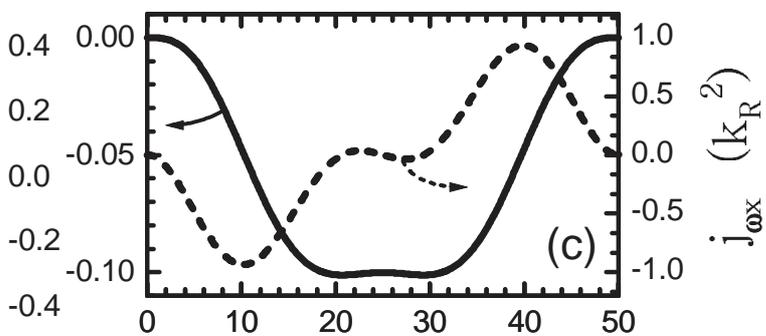
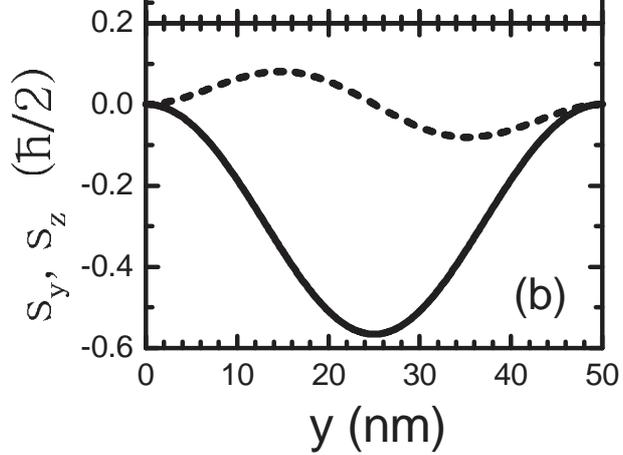
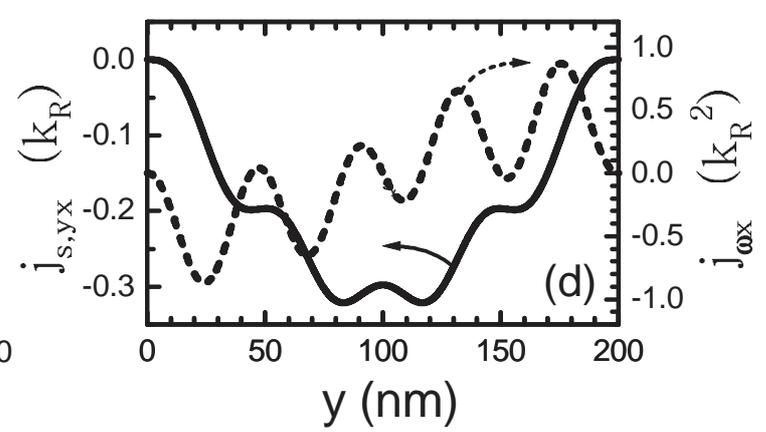
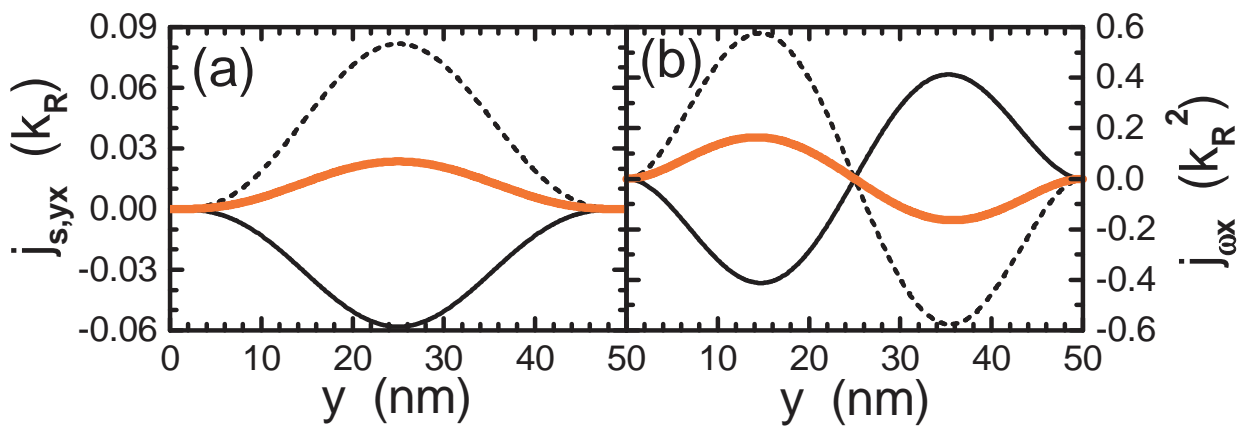
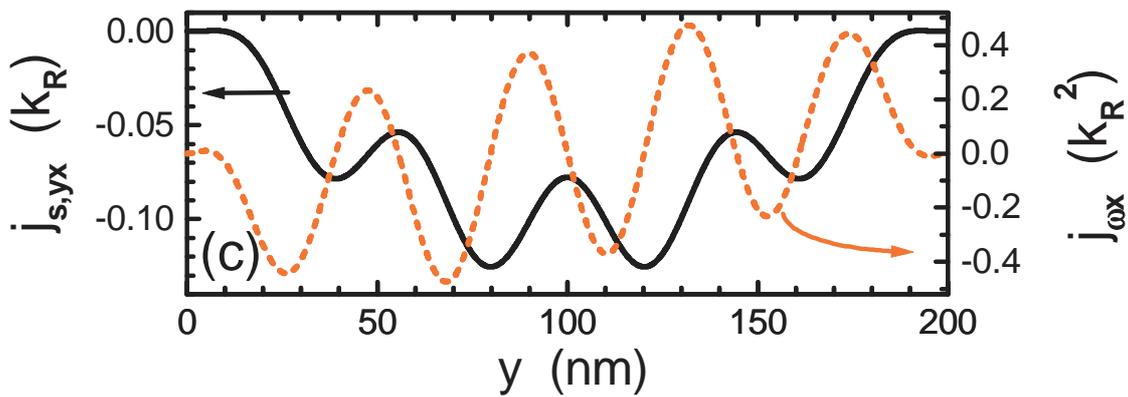
\end{document}